# $^{125}$Te spin-lattice relaxation in a candidate to Weyl semimetals WTe$_2$

A. O. Antonenko[1], E. V. Charnaya[1], A. L. Pirozerskii[1], D. Yu. Nefedov[1], M. K. Lee[2], L. J. Chang[2], J. Haase[3], S. V. Naumov[4], A. N. Domozhirova[4], V. V. Marchenkov[4,5]

[1]St. Petersburg State University, St. Petersburg, 198504 Russia

[2]National Cheng Kung University, Tainan, 70101 Taiwan

[3]Faculty of Physics and Geosciences, Leipzig University, Leipzig, 04103 Germany

[4]M.N. Mikheev Institute of Metal Physics, Ural Branch, Russian Academy of Sciences, Yekaterinburg, 620108 Russia

[5]Ural Federal University, Yekaterinburg, 620002 Russia

The tungsten ditelluride WTe$_2$ was suggested to belong to the Weyl semimetal family. We studied $^{125}$Te spin-lattice relaxation and NMR spectra in a WTe$_2$ single crystal within a large range from 28 K up to room temperature. Measurements were carried out on a Bruker Avance 500 NMR pulse spectrometer for two orientations of the crystalline $c$ axis in magnetic field, $c \parallel \vec{B}_0$ and $c \perp \vec{B}_0$. Relaxation proved to be single-exponential. The relaxation time varied depending on the sample position in magnetic field and frequency offset. The relaxation rate increased about linearly with temperature below 70 K however the dependence became nearly quadratic at higher temperatures. The relaxation rate within the total temperature range was fitted using a theoretical model developed in [41] for Weyl semimetals and assuming the decrease of the chemical potential with increasing temperature. The results obtained for $^{125}$Te spin-lattice relaxation evidence in favor of the topological nontriviality of the WTe$_2$ semimetal. The $^{125}$Te NMR spectra agreed with the occurrence of nonequivalent tellurium sites and varied insignificantly with temperature.

## 1. Introduction

The Weyl semimetals (WSM) have nontrivial topology of the electronic band structure and form a new class of 3D topological materials [1,2]. The topologies of WSM and of earlier discovered topological insulators originate from similar inverted bands due to strong spin-orbit coupling [3,4]. In topological insulators a full gap opens in bulk giving rise to metallic surface states. In WSM gapless band touching appears in the bulk Brillouin zone and topological surface states constitute the Fermi arcs [5,6]. The Weyl points in the bulk excitation spectrum always arise in pairs with opposite chirality. The WSM require the breaking of either the time-reversal symmetry or the lattice inversion symmetry. Both cases were considered theoretically [5,7-9], however, first Weyl semimetal was experimentally confirmed in a non-centrosymmetric TaAs and later in some other crystals from its family [10-18]. TaAs belongs to type-I WSM. The Fermi surface in such crystals shrinks to zero at the Weyl points as the Fermi energy is sufficiently close to the Weyl points. Another type of WSM, type-II, have touching points between electron and hole pockets [19]. Electron and hole bands constitute tilted Weyl cones with a finite density of states at the Weyl points. The tilted Weyl cones result in the violation of the Lorentz invariance. The hallmark of WSM with breaking inversion symmetry is that the total number of Weyl points must be a multiple of four.

It was claimed that the layered tungsten and molybdenum ditellurides, $WTe_2$, and $MoTe_2$, as well as their alloys $Mo_xW_{1-x}Te_2$ are type-II WSM [20-28] although the evidences for $WTe_2$ were questioned [29]. Up to now, no unambiguous confirmations of the topological nature of $WTe_2$ were obtained [2,25,29-31]. However, convincing arguments for $WTe_2$ being a type-II Weyl semimetal were recently provided with electrical transport measurements [32,33] and visualization of the Weyl nodes and Fermi arcs by scanning tunneling microscopy [26]. Besides an interest to $WTe_2$ as a likely Weyl semimetal, this layered compound may open new possibilities for materials engineering and applications including quantum computing. $WTe_2$ also demonstrated the Lifshitz transition driven by temperature [34].

While the Dirac cones in 3D topological insulators emerge only on their surfaces [35], WSM host a linear dispersion in bulk. It allows expecting that experimental techniques, which are sensitive to bulk properties, could help to find specific features in $WTe_2$, inherent to type-II Weyl semimetals. NMR line shift and shape and nuclear spin relaxation in conductors strongly depend on electron-nuclear hyperfine coupling and then can be responsive to specific linear dispersion near the Weyl points contrary to the case of the topological insulators [36-39]. Our first $^{125}$Te NMR studies in $WTe_2$ revealed a complex spectrum corresponded to non-equivalent tellurium sites [40], which did not show pronounced changes between room and low temperatures. Here we present $^{125}$Te spin-lattice relaxation and NMR spectra measurements in a $WTe_2$ single crystal within a large temperature range. We found that nuclear spin relaxation was single-exponential with a relaxation time $T_1$, which was proportional to the inverse temperature in accordance to the

Korringa law below 70 K and strongly deviated from this dependence above 70 K up to room temperature. The temperature evolution of relaxation obtained was treated assuming the nontrivial topology in WTe$_2$ and using a theoretical model developed for WSM [41,42].

**2. Sample and experiment**

The layered tungsten ditelluride single crystal was grown by the chemical vapor transport method with bromine as the transport agent. The growth process in an evacuated and sealed quartz ampoule lasted three weeks. A plate for NMR measurements was split from the grown ingot perpendicular to the crystalline *c* axis. The plate size was 0.2x3x4 mm. WTe$_2$ crystallizes in the *Td* (orthorhombic, non-centrosymmetric space group *Pmn*2$_1$) phase [43], which remains stable under temperature variation [44,45]. The X-ray diffraction at room temperature confirmed the *Td* phase and the single-crystallinity of the sample.

The WTe$_2$ structure consists of triple atom layers, Te-W-Te, stacking along the *c*-axis. The layers are bounded by weak van der Waals interactions, the W-Te bonds being covalent. The unit cell comprises four formula units with eight tellurium and four tungsten atoms (Fig. 1). The eight tellurium sites are crystallographically equivalent in pairs. When the magnetic field in NMR experiments is parallel or perpendicular to the *c*-axis, the atoms in Fig. 1 marked with apostrophe are magnetically equivalent to their counterparts without apostrophe. Totally, at the particular orientations of WTe$_2$ we have four non-equivalent tellurium sites marked with digits from 1 to 4 in Fig. 1.

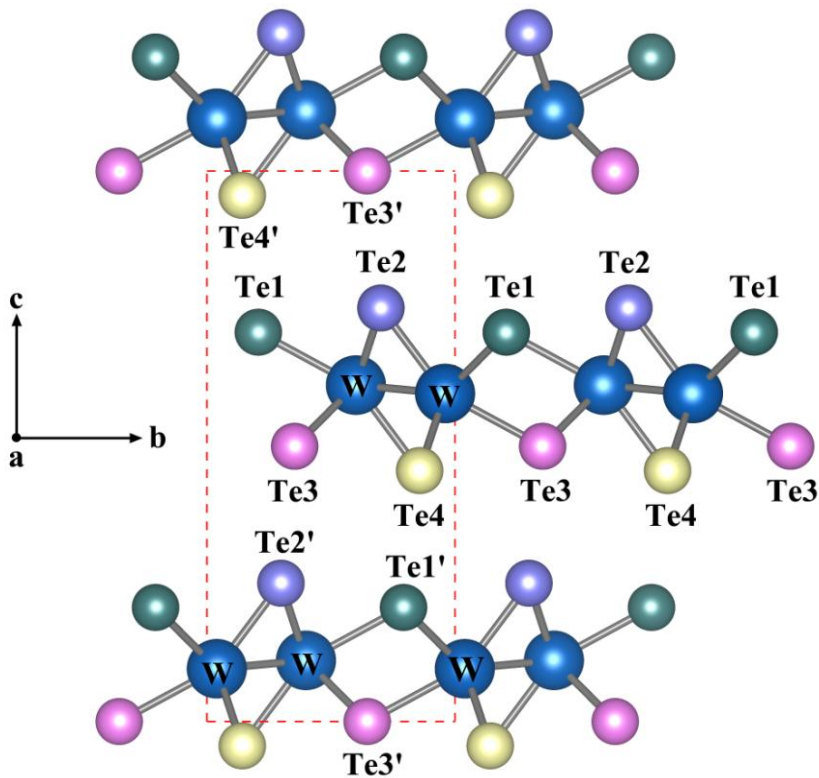

Fig. 1. Structure of *Td* WTe$_2$.

The $^{125}$Te NMR experiments were carried out using a Bruker Avance 500 NMR spectrometer in the magnetic field 11.75 T. The WTe$_2$ crystalline $c$ axis was set in parallel $c \| \vec{B}_0$ and perpendicularly $c \perp \vec{B}_0$ to the magnetic field $\vec{B}_0$. To detect the NMR signals a spin-echo pulse sequence was applied with a $\pi/2$-pulse from 4.5 to 5.5 $\mu$s depending on temperature and the sample orientation in magnetic field. Two kinds of spectra were collected. First, the intensities of spin echo at variable frequencies were found and then the envelope was plotted. Second, the echo data at a stepped offset were summed up to get variable offset cumulative spectra. The frequency of $^{125}$Te NMR was calibrated using the unified scale $\Xi$ [46]. The spin-lattice relaxation time $T_1$ was measured with saturation recovery spin echo procedure. Two and three frequency offsets within the spectra for the sample orientations in magnetic field with $c \| \vec{B}_0$ and $c \perp \vec{B}_0$, respectively, as described in Section 3, were used for relaxation measurements. The $^{125}$Te low natural abundance (~7 %) leads to weak NMR signals. Then to get a sufficient signal-to-noise ratio, the number of scans in each our experiment was larger than $2^{11}$. The measurements were performed from 28 K to room temperature with a low-temperature wide-line Bruker HPLTBB probe. Temperature stabilization was better than 2 K.

**3. Results and discussion**

The complex $^{125}$Te NMR spectra observed for both crystal orientations agree with the results presented in [40]. Fig. 2 shows the spectra at 41 and 180 K as examples. At $c \| \vec{B}_0$ the spectrum at temperatures below 70 K consists of four lines (Fig. 2(*c*)) with frequency shifts in the range 800-2000 ppm. The four lines correspond to four non-equivalent tellurium sites in the unit cell. The lines slightly broaden and overlap pairwise at higher temperatures (Fig. 2(*a*)). When $c \perp \vec{B}_0$ the $^{125}$Te NMR spectra demonstrate three lines in the whole temperature range (Fig. 2(*b,d*)), two of the four non-equivalent telluriums having similar frequency shifts. The spectra at both orientations of the sample in magnetic field do not shift significantly with changing temperature.

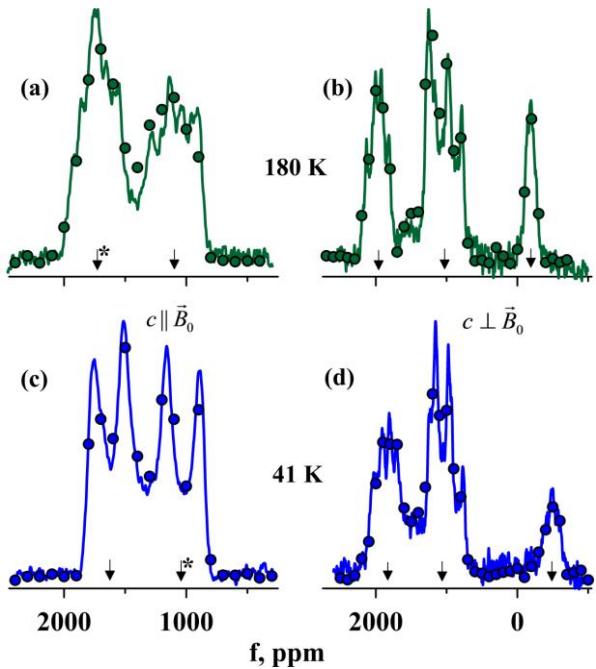

Fig. 2. $^{125}$Te NMR spectra at 180 K (upper panels) and 41 K (lower panels). The panels (a,c) and (b,d) show spectra at $c \| \vec{B}_0$ and $c \perp \vec{B}_0$, respectively. Solid lines - variable offset cumulative spectra. Circles – echo signal intensities.

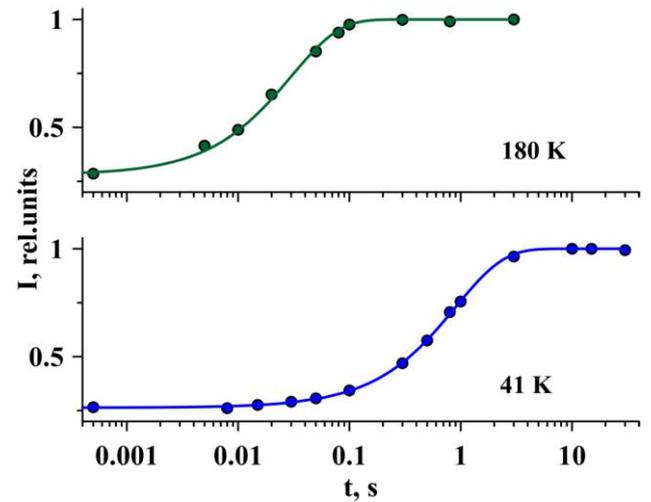

Fig. 3. Magnetization recovery curves obtained for $c \| \vec{B}_0$ at frequency offsets marked with asterisks in Fig. 2 (a) (upper panel) and (c) (lower panel). The solid lines show the single-exponential recovery.

Because the spectra were very broad we measured relaxation for several frequency ranges separately. For this we chose two and three offsets of frequency at $c \| \vec{B}_0$ and $c \perp \vec{B}_0$, respectively, as shown by arrows in Fig. 2. The magnetization recovery proved to be single-exponential at different temperatures and therefore could be described by relaxation times $T_1$ (Fig. 3). The relaxation times varied with the frequency offset and sample orientation in magnetic field. The temperature dependences of the relaxation rates $1/T_1$ are shown in Fig. 4. For both orientations of the WTe$_2$ sample in magnetic field the relaxation time at the lowest temperature in our experiments was of the order of a second. The relaxation rates gradually increase with increasing temperature. Below about 70 K the relaxation rates rises near linearly with temperature. However, their rising becomes stronger at higher temperature yielding about quadratic temperature dependences.

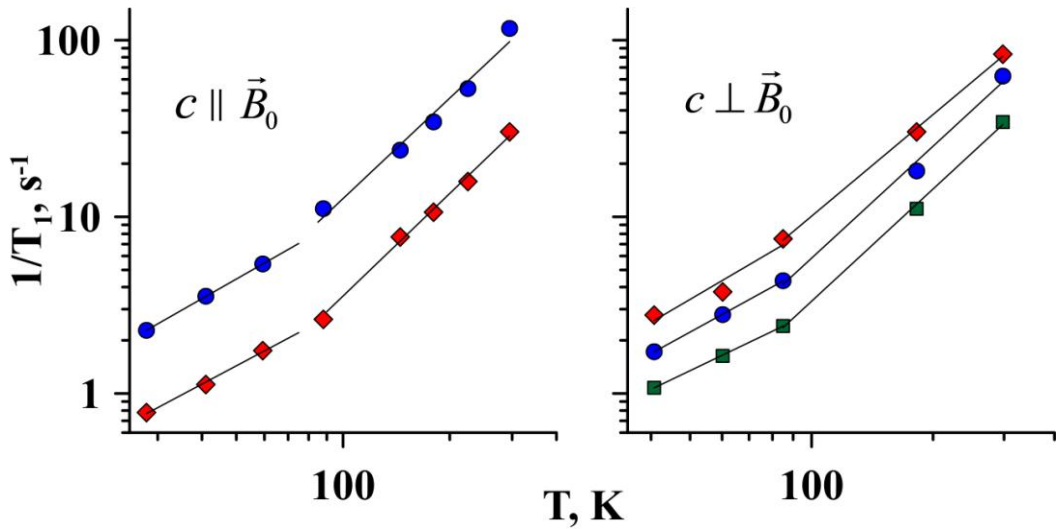

Fig. 4. $^{125}$Te relaxation rates obtained for $c \| \vec{B}_0$ (left panel) and $c \perp \vec{B}_0$ (right panel). Left panel: circles and diamonds correspond to larger and smaller frequency offsets shown in Fig. 2, respectively. Right panel: circles, squares, and diamonds correspond to larger, central, and smaller frequency offsets shown in Fig. 2, respectively. Straight lines show power fits.

The linear rise of the relaxation rate with temperature is typical for metals, in which the main contribution to nuclear spin-lattice relaxation is due to the hyperfine coupling with conduction electrons [47] especially if the nuclear spin is ½ as for the $^{125}$Te isotope and quadrupole relaxation by virtue of spin-phonon coupling [48] is absent. For semiconductors and semimetals with trivial topology the temperature dependence of the nuclear spin relaxation rate can be distorted and enhanced owing to thermal activation processes and evolution of the electronic band structure [49-51]. For instance, the thermal activation processes were involved to treat the strong temperature dependence of Te spin relaxation in the PbTe semiconductor [51]. The quadratic temperature dependence of tellurium spin relaxation in WTe$_2$ might result from changes of the band structure induced by temperature [34,52,53].

On the other hand, if we assume that WTe$_2$ is a type-II Weyl semimetal, then the bend on the temperature dependence of the $^{125}$Te spin-lattice relaxation rate can be caused by the impact of the Weyl fermions [41,42]. The band-structure calculations for WTe$_2$ suggested 8 Weyl nodes in the bulk Brillouin zone [19]. It was found in [41] that the presence of the Weyl points in WSM change entirely the hyperfine coupling. In conductors with trivial topology the hyperfine coupling varies feebly with temperature yielding, in particular, a weak temperature dependence of the Knight shift [47]. In WSM the opposite is true: the hyperfine coupling depends strongly on the chemical potential and temperature. The hyperfine coupling is dominated by the orbital part, which differs significantly from that in conductors with trivial topology. It diverges upon approaching the Weyl point and overwhelms the conventional, spin-dipole and Fermi contact, interactions. The orbital part of the hyperfine coupling provides the largest contribution to nuclear spin relaxation [41]. Similar results were obtained for Dirac semimetals in [54]. It was found in [41] that the dominant term in the nuclear spin-lattice relaxation rate $1/T_1$ in WSM is given by

$$\frac{1}{T_1} = \frac{\pi \mu_0^2 \gamma_n^2}{4 v_F (2\pi)^6} \int_{-\infty}^{\infty} dk \frac{(k e v_F)^2 F(|k|/k_0)}{\cosh^2[(\hbar v_F k - \mu)/2 k_B T]}, \qquad (1)$$

where $k_B$ and $\hbar$ are the Boltzmann and reduced Planck constants, respectively, $k_0 = \omega_0/v_F$, $\omega_0$ Larmor frequency, $v_F$ Fermi velocity, $\mu_0$ vacuum permeability, $\gamma_n$ nuclear gyromagnetic ratio, $e$ electronic charge, and $\mu$ chemical potential. Relationship (1) can be rewritten as

$$\frac{1}{T_1} = \frac{4 \mu_0^2 \gamma_n^2 e^2 k_B^3}{3 h^3} \frac{T^3}{v_F^2} (t_1 + t_2 + t_3), \qquad (2)$$

where $t_1 = \frac{1}{2}\left(\frac{\mu}{k_B T}\right)^2 \ln\left(\frac{4 k_B T}{\hbar \omega_0}\right)$, $t_2 = \frac{\pi^2}{6} \ln\left(\frac{4 k_B T}{\hbar \omega_0}\right)$, $t_3 = \frac{1}{2} \int_{-\infty}^{+\infty} \frac{x^2 \ln(x^2)}{\cosh^2\left(x - \frac{\mu}{2 k_B T}\right)} dx$.

At high temperature when $\mu \ll k_B T$ the contribution of $t_2$ dominates and

$$\frac{1}{T_1} \propto T^3 \ln\left(\frac{4 k_B T}{\hbar \omega_0}\right). \qquad (3)$$

At low temperature when $\mu \gg k_B T$ the temperature dependence of the relaxation rate is mainly determined by $t_1$, which yields

$$\frac{1}{T_1} \propto T \ln\left(\frac{2\mu}{\hbar\omega_0}\right). \tag{4}$$

According to (4), at low temperatures the relaxation rate rises proportionally to temperature as in metals if the chemical potential changes weakly. It agrees with our relaxation measurements up to 70 K. The temperature dependence of the relaxation rate at higher temperatures is weaker in WTe$_2$ than predicted by (3). The spin-lattice relaxation rate, which rose quadratically with temperature, was also observed for $^{113}$Cd in the Cd$_3$As$_2$ Dirac semimetal [55] within a range from 225 to 300 K. On the other hand, a dependence $1/T_1 \propto T^3$ was found for $^{181}$Ta by nuclear quadrupole resonance in the TaP Weyl semimetal above ~30 K [56] and was treated later in [42] on the basis of the relationship (1) implying a temperature-dependent chemical potential. We used the theory developed in [41] to fit the variation of $^{125}$Te $1/T_1$ with temperature obtained for the WTe$_2$ sample orientation $c \| \vec{B}_0$ at a larger frequency offset indicated by an asterisk in Fig. 2(a) and presented on the left panel in Fig. 4 (circles).

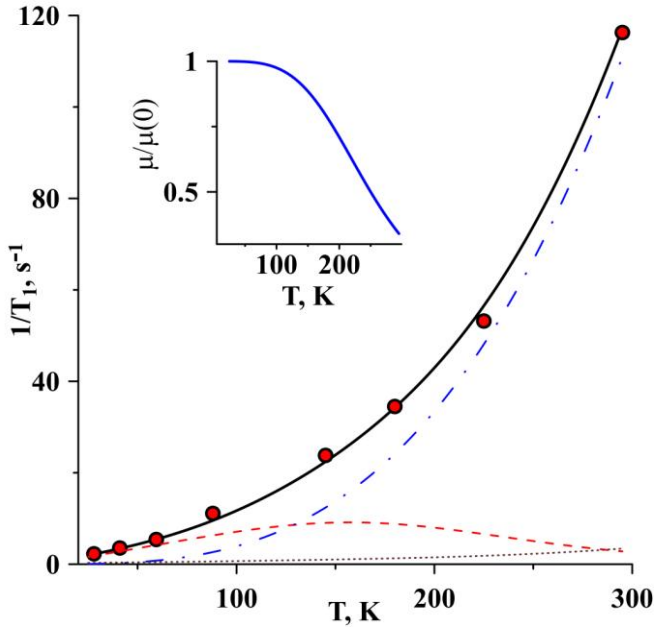

Fig. 5. $^{125}$Te spin-lattice relaxation rate obtained for $c \| \vec{B}_0$ at a larger frequency offset indicated by an asterisk in Fig. 2(a). Circles – experimental data. The solid, dashed, dash-and-dot, and dotted lines show the fit using Relationship (2) and contributions proportional to $t_1$, $t_2$, and $t_3$, respectively. The temperature dependence of the chemical potential is shown in the inset.

The result of fitting is shown in Fig. 5. To get a good agreement with the experimental data we assumed following [42] that the chemical potential decreased with increasing temperature. The temperature dependence of the chemical potential is demonstrated in the inset to Fig. 5. One can see in Fig. 5 that in the intermediate temperature range where the contribution of $t_1$ is still essential, the decrease in the chemical potential weakens the temperature dependence of the relaxation rate and makes it close to $T^2$. The fitting parameters, which correspond to Fig. 5, are $\mu(T=0) = 250$ K and $v_F = 127$ m/s. Note that the estimate for the Fermi velocity is much smaller than the usual values for semimetals of the order of $10^5$ m/s.

The shifts of the NMR resonance frequency for the $^{125}$Te nuclei in WTe$_2$ can be caused by hyperfine coupling with charge carriers (Knight shift) and by chemical shift caused by coupling with bonding electrons. Contrary to many lighter nuclei, the chemical shift for tellurium can be high, of the order of 4000 ppm [57,58]. When the chemical shift is large, it is difficult to separate it from the Knight shift. In [37,38] the Knight and chemical shifts for $^{125}$Te in the Bi$_2$Te$_3$ topological insulator were distinguished using their different temperature dependences. The Knight shift in metals is known to depend weakly on temperature in contrast to semiconductors and semimetals. For semimetals the temperature dependence of the Knight shift can be pronounced even below room temperature because of the low Fermi energy (and associated fast variation of the chemical potential) [42] and transformation of the electron and hole pockets. For the particular case of WTe$_2$ the Knight shift should be additionally affected by pronounced changes in the electron and hole bands upon increasing temperature, which led to the Lifshitz transition [34,52,53]. It was shown in [34] that the band transformation is associated with the complete disappearance of the hole pockets above ~160 K. On the contrary, the chemical shift varies feebly with temperature except for the ranges of the structural phase transitions. The rather weak change of the frequency shift of the $^{125}$Te NMR spectra in WTe$_2$, which we observed, can be likely due to domination of the chemical shift over the Knight shift. Then we can hardly expect that the Korringa relation [47]

$$T_1 K_S^2 T = \frac{\hbar}{4\pi k_B} \left( \frac{g\mu_B}{\hbar \gamma_n} \right)^2 \qquad (5)$$

is satisfied between the $^{125}$Te relaxation times and line positions shown in Fig. 2. Here $g$ is the Landé factor, $\mu_B$ Bohr magneton, $K_S$ the Knight shift. Nevertheless, for both frequency offsets at the orientation $c \| \vec{B}_0$ the product $T_1 T (\Delta f)^2$ ($\Delta f$ is the experimental shift) below 70 K is very close to 2.5 $10^{-5}$ s·K in the Korringa relation for the $^{125}$Te isotope (about 3 $10^{-5}$ s·K and 3.8 $10^{-5}$ s·K for the higher and lower frequency offsets, respectively). At higher temperatures this correspondence worsens, the product decreasing to about $10^{-5}$ s·K at room temperature. At the orientation $c \perp \vec{B}_0$ the Korringa relation is sufficiently satisfied at low temperatures only for a central line with a shift ~1000 ppm, the product $T_1 T (\Delta f)^2$ being equal to 3.8 $10^{-5}$ s·K approximately.

**Conclusions.** NMR measurements showed that $^{125}$Te spin-lattice relaxation in the WTe$_2$ semimetal single crystal is single-exponential within a range from 28 K up to room temperature. The relaxation rates $1/T_1$ found at two crystal orientations in magnetic field demonstrated specific dependences on temperature. Below 70 K the relaxation rates increased about linearly with temperature in a way typical for conductors with trivial topology. Above 70 K the temperature dependences intensified and became close to quadratic. The relaxation rates within the total temperature range allowed the treatment using a theoretical model developed in [41] for WSM and implying a decrease of the chemical potential with increasing temperature. The particular temperature dependence of the $^{125}$Te spin-lattice relaxation rate provided additional arguments in favor of a statement that WTe$_2$ belonged to the WSM family.

Acknowledgements. The authors acknowledge the financial support from RFBR, grants 19-57-52001 and 19-07-00028. The work was partly supported by Minobrnauki of Russia ("Spin" No. AAAA-A18-118020290104-2) and Russian Government (contract No. 02.A03.21.0006). Measurements were partly carried out using the equipment of the Research park of St. Petersburg State University.